\documentclass[aps,pra,reprint,superscriptaddress,showpacs]{revtex4-1}

\usepackage[dvips]{graphicx}  
\usepackage{amsmath,amssymb,bm}
\usepackage{color}

\begin{document}


\title{Unpolarized, incoherent repumping light for prevention of dark states in a trapped and laser-cooled single ion}


\author{T.\ Lindvall}
\email[]{thomas.lindvall@mikes.fi}
\affiliation{Centre for Metrology and Accreditation (MIKES), P.O.\ Box 9, FI-02151 Espoo, Finland}
\author{T.\ Fordell}
\affiliation{Centre for Metrology and Accreditation (MIKES), P.O.\ Box 9, FI-02151 Espoo, Finland}
\author{I.\ Tittonen}
\affiliation{Department of Micro- and Nanosciences, Aalto University, P.O.\ Box 13500, FI-00076 Aalto, Finland}
\author{M.\ Merimaa}
\affiliation{Centre for Metrology and Accreditation (MIKES), P.O.\ Box 9, FI-02151 Espoo, Finland}


\date{\today}

\begin{abstract}
Many ion species commonly used for laser-cooled ion-trapping studies have a low-lying metastable $^2D_{3/2}$ state that can become populated due to spontaneous emission from the $^2P_{1/2}$ excited state. This requires a repumper laser to maintain the ion in the Doppler cooling cycle. Typically, the $^2D_{3/2}$ state, or some of its hyperfine components if the ion has nuclear spin, has a higher multiplicity than the upper state of the repumping transition. This can lead to dark states, which have to be destabilized by an external magnetic field or by modulating the polarization of the repumper laser.
We propose using unpolarized, incoherent amplified spontaneous emission (ASE) to drive the repumping transition.
An ASE source offers several advantages compared to a laser. It prevents the buildup of dark states without external polarization modulation even in zero magnetic field, it can drive multiple hyperfine transitions simultaneously, and it requires no frequency stabilization.
These features make it very compact and robust, which is essential for the development of practical, transportable optical ion clocks. We construct a theoretical model for the ASE radiation, including the possibility of the source being partially polarized. Using $^{88}\mathrm{Sr}^+$ as an example, the performance of the ASE source compared to a single-mode laser is analyzed by numerically solving the eight-level density-matrix equations for the involved energy levels. Finally, a reduced three-level system is derived, yielding a simple formula for the excited-state population and scattering rate, which can be used to optimize the experimental parameters.
The required ASE power spectral density can be obtained with current technology.
\end{abstract}


\maketitle


\section{Introduction \label{sec:introduction}}

Optical frequency standards based on single trapped ions already have accuracies better than that of the primary cesium fountain standards \cite{Rosenband2008a,Chou2010a,Huntemann2012a,Madej2012b}, and the International Committee for Weights and Measures (CIPM) recommends quadrupole transitions in three ions ($^{88}$Sr$^+$, $^{199}$Hg$^+$, and $^{171}$Yb$^+$) as secondary representations of the second \cite{CIPM2006a}. Amendments to this list have been proposed by the joint consultative working group on frequency standards.
To characterize the optical clocks at the lowest-obtainable uncertainty level, local comparisons between clocks from different institutes are needed. This requires at least one of the clocks to be transportable. Compared to standard laboratory setups, this places stringent conditions regarding compactness and robustness on all parts of the clock, including the light sources.

\begin{figure}[b]
\includegraphics[width=.8\columnwidth]{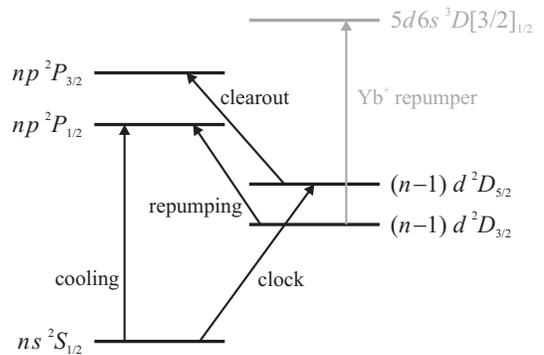}%
\caption{\label{fig:energy-levels} Lowest energy levels and relevant transitions for ions Ca$^+$ ($n=4$), Sr$^+$ ($n=5$), and Ba$^+$ ($n=6$). Items in gray apply to Yb$^+$ only ($n=6$).}
\end{figure}

Many ions commonly used for ion clocks have an alkali-metal--like atomic structure. The analysis in this paper focuses on the $^{88}\mathrm{Sr}^+$ ion, but the applicability to other ions will be discussed in Sec.~\ref{sec:ions}. The trapped ion is Doppler cooled and fluorescence detected using a cooling laser driving the $^2S_{1/2}$-$\,^2P_{1/2}$ transition (see Fig.~\ref{fig:energy-levels}). The $^2S_{1/2}$-$\,^2D_{5/2}$ quadrupole transition is used as the reference or clock transition. The $^2P_{1/2}$ excited state has a finite probability of decaying to the low-lying metastable $^2D_{3/2}$ state and a repumper laser tuned to the $^2D_{3/2}$-$\,^2P_{1/2}$ transition is required to maintain the ion in the cooling cycle. As the multiplicity of the $^2D_{3/2}$ state is higher than that of the $^2P_{1/2}$ state, a single-mode repumper laser will create dark states within the $^2D_{3/2}$ sublevels if the external magnetic field is weak \cite{Berkeland2002a,Lindvall2012a}, which is usually required in ion clocks. This causes the scattering rate to drop to nearly zero, which inhibits both the cooling and the detection of the ion. Dark states can be prevented by modulating the polarization of the repumper \cite{Barwood1998a,Berkeland1998b}, e.g., by using an electro-optic modulator (EOM). In addition, the laser must be frequency stabilized. Alternatively, one can use a multimode fiber laser and overlap two beams with different polarizations \cite{Sinclair2001a}, which results in a fluctuating net polarization.

We propose using unpolarized, incoherent amplified spontaneous emission (ASE) to drive the repumping transition.
The ASE source offers several advantages compared to a laser repumper. It prevents the buildup of dark states without polarization modulation using a single beam even in a zero-magnetic-field environment.
Due to the broad bandwidth of the source, it requires no frequency stabilization and, in ions with nuclear spin, it can excite multiple hyperfine transitions simultaneously.
These features, and the fact that the ASE source can be fiber based, make it very compact and robust, which is essential for the development of practical, transportable optical clocks. A similar ASE source at the $^2D_{5/2}$-$\,^2P_{3/2}$ transition wavelength could be used to replace the so-called clearout laser, whose purpose is to shorten the measurement cycle by emptying the metastable $^2D_{5/2}$ state after a quadrupole transition has occurred.

In the context of ion traps, incoherent light sources in the form of UV light-emitting diodes (LEDs) have previously been used for photoionization of calcium \cite{Lucas2004a,Tanaka2005a,Schuck2010a} and barium \cite{Wang2011b}.

This paper is organized as follows: In Sec.~\ref{sec:ASEmodel}, we consider two well-known models for stochastic fields and construct a model with adjustable intensity correlations for the ASE field. In Sec.~\ref{sec:8-level}, we solve the eight-level density-matrix equations of the ion numerically and compare the different field models with each other and with a single-mode laser repumper. The effect of a partially polarized field is analyzed in Sec.~\ref{sec:PP field}, and in Sec.~\ref{sec:3-level} we derive a reduced three-level system that can be solved exactly. In Sec.~\ref{sec:experimental} some experimental questions are addressed and in Sec.~\ref{sec:ions} we discuss which other ion species this concept can be applied to. Finally, the results are summarized.

\section{Amplified spontaneous emission field model \label{sec:ASEmodel}}

The electric field of the ASE light at the position of the trapped ion can be represented by the sum of two components with orthogonal polarizations ($\mathbf{u}_1\cdot\mathbf{u}_2=0$):
\begin{equation} \label{eq:unpol}
\mathbf{E}_\mathrm{ASE}(t) = \frac{1}{2^{3/2}} [\mathbf{u}_1 \mathcal{E}_1(t) + \mathbf{u}_2 \mathcal{E}_2(t)] e^{-i\omega_\mathrm{ASE} t} + \text{c.c.},
\end{equation}
where $\omega_\mathrm{ASE}$ is the center frequency of the field and $\text{c.c.}$ stands for complex conjugate.
If the light is unpolarized, the components $\mathcal{E}_1(t)$ and $\mathcal{E}_2(t)$ are completely uncorrelated stochastic variables and their intensities are equal \cite{Lahiri2012a}.

We use Ornstein-Uhlenbeck (OU) processes \cite{Uhlenbeck1930a} to model the stochastic variables, which is convenient as there are exact updating formulas for both the OU process and its time integral \cite{Gillespie1996a}. The OU process $X(t)$ obeys the Langevin equation
\begin{equation}
\frac{d}{dt} X(t) = -\frac{1}{\tau_\mathrm{r}} X(t) + c^{1/2}\Gamma(t),
\end{equation}
where $\tau_\mathrm{r}$ and $c$ are positive constants, traditionally referred to as the relaxation time and diffusion constant, respectively, and $\Gamma(t)$ is delta-correlated Gaussian white noise fulfilling $\langle \Gamma(t) \Gamma(t')\rangle = \delta(t-t')$. The correlation function for $X(t)$ is given by
\begin{equation} \label{eq:OUcorr}
\langle X(t) X(t+\tau)\rangle = \frac{c\tau_\mathrm{r}}{2} e^{-|\tau|/\tau_\mathrm{r}}.
\end{equation}

Two common types of stochastic fields are the phase-diffusion (PD) field and the chaotic field \cite{Georges1979a}. The phase-diffusion field,
\begin{equation} \label{eq:E_PD}
\mathcal{E}_{\text{PD},j}(t) = \mathcal{E}_0 e^{-i \phi_j(t)},
\end{equation}
has a constant amplitude and a fluctuating phase and is typically used to describe a single-mode laser of finite bandwidth \cite{Dixit1980a,Stenholm:FLS,Camparo1999a,Blushs2004a}. In this model, the time derivatives of the phases, i.e., the frequency fluctuations, fulfill the correlation function (\ref{eq:OUcorr}),
\begin{equation}
\langle \dot{\phi}_i(t) \dot{\phi}_j(t+\tau)\rangle = \frac{b\beta}{2} e^{-\beta|\tau|} \delta_{ij},
\end{equation}
with $\beta = \tau_\mathrm{r}^{-1}$ and $b = c\tau_\mathrm{r}^2$. The delta function indicates that the two polarization components are uncorrelated.
The (normalized) first-order correlation function of the phase-diffusion field is \cite{Dixit1980a}
\begin{eqnarray}
g^{(1)}_{\text{PD},ij}(\tau) &=& \frac{\langle \mathcal{E}_{\text{PD},i}^*(t) \mathcal{E}_{\text{PD},j}(t+\tau)\rangle}{\langle \mathcal{E}_{\text{PD},i}^*(t) \mathcal{E}_{\text{PD},j}(t)\rangle} = \langle e^{i[\phi_i(t)-\phi_j(t+\tau)]}\rangle \nonumber \\
&=& \exp \left\{-\frac{b}{2} \left[|\tau|-\frac{1}{\beta}\left(1-e^{-\beta|\tau|}\right)\right]\right\} \delta_{ij}.
\end{eqnarray}
This corresponds to a lineshape that is essentially Lorentzian within $\pm\beta$ from line center and falls off as a Gaussian outside this region. In the limit $\beta\gg b$, the correlation function becomes $g^{(1)}_{\text{PD},ij}(\tau) \approx e^{-b|\tau| /2} \delta_{ij}$, corresponding to a Lorentzian line with a full width at half maximum (FWHM) of $b$. In this case the OU process approximates a Gaussian white noise (Wiener) process. As the amplitude of the PD fields is constant, the second-order correlation function is $g^{(2)}_{\text{PD},ij}(\tau) = 1$.

We now turn to the chaotic field,
\begin{equation} \label{eq:E_ch}
\mathcal{E}_{\text{ch},j}(t) = 2^{-1/2}[\mathcal{E}_j^\mathrm{r}(t) + i \mathcal{E}_j^\mathrm{i}(t)],
\end{equation}
where the real and imaginary parts are independent Gaussian stochastic variables. It has both amplitude and phase fluctuations and can be used to describe a multimode laser field with a large number of uncorrelated modes \cite{Georges1979a}.
Now the electric field fulfills Eq.~(\ref{eq:OUcorr}),
\begin{equation}
\langle \mathcal{E}_i^k(t) \mathcal{E}_j^l(t+\tau)\rangle =  \mathcal{E}_0^2 e^{-b|\tau|/2} \delta_{ij} \delta_{kl},
\end{equation}
where $b=2/\tau_\mathrm{r}$ and $\mathcal{E}_0^2=c\tau_\mathrm{r}/2$. This gives the first-order correlation function $g^{(1)}_{\text{ch},ij}(\tau) = e^{-b|\tau| /2} \delta_{ij}$, identical to that of the PD model in the $\beta\gg b$ limit. Thus also the spectrum of the chaotic field is a Lorentzian with a FWHM of $b$.
For a real Gaussian process, the first-order correlation function contains all the information about the higher-order correlation functions [see, e.g., Eq.~(2.1--8) in Ref.~\cite{Mandel-OCCO}] and the second-order correlation function becomes $g^{(2)}_{\text{ch},ij}(\tau) = 1+e^{-b|\tau|}\delta_{ij}$, which shows photon bunching.

The second-order correlation properties of the polarized ASE from a superluminescent light-emitting diode (SLED) have recently been studied \cite{Blazek2011a}. It was shown that $g^{(2)}_\text{ASE}(\tau)$ depends on the pump current of the SLED: at low currents the SLED emits pure chaotic light, $g^{(2)}_\text{ASE}(0)=2$, but as the current increases, the value of $g^{(2)}_\text{ASE}(0)$ decreases, being $1.3$ at the highest current. We expect other ASE sources to show a similar dependence on the pump and have therefore constructed a field model with adjustable intensity correlations. Expressing the chaotic field as $\mathcal{E}_{\text{ch},j}(t) = 2^{-1/2}\{[\mathcal{E}_j^\mathrm{r}(t)]^2 + [\mathcal{E}_j^\mathrm{i}(t)]^2\}^{1/2} e^{i \arg{[\mathcal{E}_j^\mathrm{r}(t) + i \mathcal{E}_j^\mathrm{i}(t)]}} \equiv |\mathcal{E}_{\text{ch},j}(t)| e^{i\arg{\mathcal{E}_{\text{ch},j}(t)}}$, we write the ASE field as
\begin{equation} \label{eq:E_ASE}
\mathcal{E}_{\text{ASE},j}(t) = \left[\mathcal{E}_0^2 \cos^2{\alpha} + |\mathcal{E}_{\text{ch},j}(t)|^2 \sin^2{\alpha} \right]^{1/2} e^{i\arg{\mathcal{E}_{\text{ch},j}(t)}}.
\end{equation}
We have numerically verified that the first-order correlation function of also this field is $g^{(1)}_{\text{ASE},ij}(\tau) = e^{-b|\tau| /2} \delta_{ij}$. The second-order correlation function can be evaluated in closed form and is $g^{(2)}_{\text{ASE},ij}(\tau) = 1+ \sin^4{\alpha}\,e^{-b|\tau|} \delta_{ij}$. Hence the spectrum is a Lorentzian with a FWHM of $b$, while the intensity correlations can be adjusted using the parameter $\alpha$.

Note that while Eq.~(\ref{eq:E_ASE}) is identical to the chaotic field $\mathcal{E}_{\text{ch},j}(t)$ for $\alpha = \pi/2$, it will not obey the same statistics as the PD field $\mathcal{E}_{\text{PD},j}(t)$ when $\alpha = 0$. It should rather be interpreted as a chaotic field with suppressed intensity correlations. This is not considered a problem, as the ASE source will never fully exhibit a single-mode-laser--like behavior. In addition, the repumping efficiency does not depend on the actual statistics, as will be shown in Sec.~\ref{sec:8-level}.

The correlation functions given so far have been for the separate polarization components, as indicated by the subscript $ij$. The first-order correlation function of the total, unpolarized field (\ref{eq:unpol}) only differs by the factor $e^{-i\omega_\mathrm{ASE} \tau}$. However, the second-order correlation function for an unpolarized light beam (for a measurement that does not distinguish the two polarizations) is given by $g^{(2)}(\tau) = [1+g_{ii}^{(2)}(\tau)]/2$ \cite{Loudon-QTL}, where $g_{ii}^{(2)}(\tau)$ is the second-order correlation function of the polarization components. This means that the photon bunching is reduced by a factor of two for unpolarized light.

\begin{figure}[h]
\includegraphics[width=1\columnwidth]{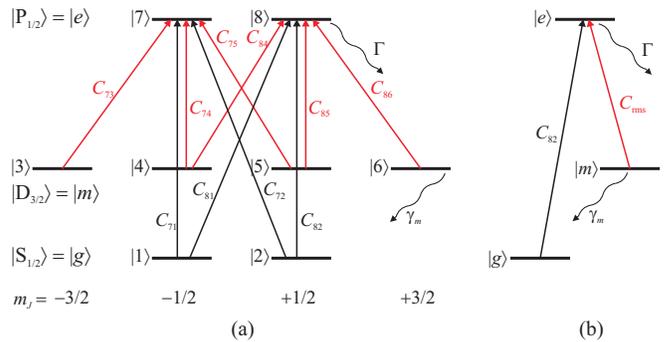}%
\caption{\label{fig:8_3-level}(Color online) (a) Eight-level system. Fine structure state designations are given at the left and magnetic quantum numbers at the bottom. The sublevels are labeled consecutively and optical dipole transitions are shown as arrows together with their corresponding relative transition amplitudes $C_{ij}$. The spontaneous decay rates are shown next to the wavy arrows. (b) Reduced three-level system obtained from the eight-level system in (a) as described in Sec.~\ref{sec:3-level}.}
\end{figure}

\section{Eight-level simulations \label{sec:8-level}}

We want to demonstrate that the ASE repumper prevents the buildup of dark states in the metastable $^2D_{3/2}$ state and between the metastable state and the $^2S_{1/2}$ ground state. We start by numerically solving the density-matrix equations of the eight-level system in Fig.~\ref{fig:8_3-level}(a) using the formalism described in our previous work \cite{Lindvall2012a}, with the repumper described by Eq.~(\ref{eq:unpol}) and either Eq.~(\ref{eq:E_PD}) or Eq.~(\ref{eq:E_ASE}) [the chaotic case is obtained from Eq.~(\ref{eq:E_ASE}) with $\alpha = \pi/2$]. All numerical results are for $^{88}\mathrm{Sr}^+$.

We are interested in the (near-)zero-magnetic-field situation, so we choose the quantization axis $\mathbf{u}_z$ along the linear polarization of the cooling laser that travels in the $\mathbf{u}_y$ direction. The ASE field is assumed to propagate in the same direction and a natural choice of the ASE polarization components is then $\mathbf{u}_1=\mathbf{u}_x$ and $\mathbf{u}_2=\mathbf{u}_z$. For the cooling laser we use the (two-level) Rabi frequency $\Omega_\mathrm{c}/\Gamma=1$ (for $^{88}\mathrm{Sr}^+$ this corresponds to the intensity $I_\mathrm{c}=39.8\;\text{mW}\text{cm}^{-2}$), which we have shown to give a high scattering rate with only minor power broadening \cite{Lindvall2012a}, and the detuning $\delta_\mathrm{c}/\Gamma=-0.5$, which minimizes the temperature achievable with Doppler cooling. Here  $\Gamma$ is the spontaneous decay rate of the $^2P_{1/2}$ excited state ($\Gamma/(2\pi) = 21.6$\;MHz for $^{88}\mathrm{Sr}^+$).

As the ASE field is a stochastic variable, this is true also for the repumper Rabi frequency. Therefore, we use its rms value to describe the strength of the field, $\Omega_\mathrm{r} \equiv [\langle |\Omega_\mathrm{r}(t)|^2\rangle_t]^{1/2}$, where $\langle \; \rangle_t$ refers to a time average. For a laser repumper with a linewidth $\ll \Gamma$, it is desirable to have a repumper Rabi frequency $\Omega_{\mathrm{r},\text{laser}} \approx \Omega_\mathrm{c} = \Gamma$ in order to maximize the scattering rate \cite{Lindvall2012a}. The effective intensity per bandwidth is thus proportional to $\Omega_{\mathrm{r},\text{laser}}^2/\Gamma = \Gamma$. For the ASE source, with a bandwidth $b\gg \Gamma$ and a Lorentzian lineshape, to have the same effective intensity per bandwidth, we require $\Omega_\mathrm{r}^2/b = \Gamma$, i.e., the Rabi frequency should scale as $\Omega_\mathrm{r}/\Gamma= (b/\Gamma)^{1/2}$.

Using the two-level Rabi frequency $\Omega = (\mu/\hbar) \sqrt{2I_0/\varepsilon_0 c}$ \cite{Lindvall2012a}, the peak intensity versus total power of a Gaussian beam, $I_0=2P_\text{tot}/(\pi w^2)$, where $w$ is the beam waist, and the peak power spectral density (PSD) of a Lorentzian line, $\delta P/\delta \omega = 2P_\text{tot}/(\pi b)$, we can relate the parameters used in the calculations to the power per unit wavelength of the ASE source:
\begin{equation}
\frac{\delta P}{\delta \lambda} = \pi \varepsilon_0 \Gamma \left(\frac{\hbar c}{\mu} \right)^2 \left(\frac{w}{\lambda} \right)^2 \left( \frac{\Omega_\mathrm{r}}{\Gamma}\right)^2 \frac{\Gamma}{b}.
\end{equation}
For $^{88}\mathrm{Sr}^+$ this is
\begin{equation}
\frac{\delta P}{\delta \lambda} \approx 5.4 \frac{\mu\mathrm{W}}{\mathrm{nm}} \left(\frac{w}{\lambda} \right)^2 \left( \frac{\Omega_\mathrm{r}}{\Gamma}\right)^2 \frac{\Gamma}{b}.
\end{equation}

As it is common to use a bandpass filter at the cooling-laser wavelength for fluorescence detection, we define the scattering rate as $\Gamma_\mathrm{sc} = A_g \Gamma (\rho_{77}+\rho_{88})$, where $A_g$ is the decay probability to the ground state.

\begin{figure}[t]
\includegraphics[width=1\columnwidth]{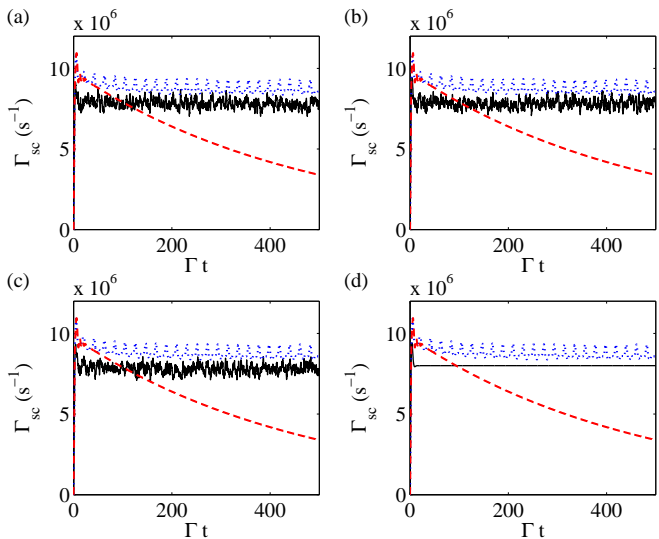}%
\caption{\label{fig:t-plot}(Color online) Scattering rate as a function of time for (a) a PD repumper, (b) a chaotic repumper, (c) an ASE repumper with $g^{(2)}_{\text{ASE},ij}(0) = 1.5$, and (d) the three-level solution from Sec.~\ref{sec:3-level} [solid (black) curves]. The curves in panels (a--c) are averages over 50 simulations. For comparison, the scattering rates for a polarization-modulated laser repumper [dotted (blue) curves] and for a laser repumper with stationary polarization [dashed (red) curves] are shown in each panel.}
\end{figure}

As the density matrix describes ensemble averages, but each numerical simulation corresponds to a particular realization of the stochastic field, we should average the results over a large number of simulations. Figure~\ref{fig:t-plot} shows the scattering rate for small values of $t$ for a repumper with $b/\Gamma=100$ (bandwidth approximately 2.2\;GHz or 0.0086\;nm) and $\Omega_\mathrm{r}/\Gamma= (b/\Gamma)^{1/2} = 10$ described by [Fig.~3(a)] the PD, [Fig.~3(b)] the chaotic, and [Fig.~3(c)] the ASE models [solid (black) curves]. The curves are averages over 50 simulations, so there are some fluctuations left that would vanish for a high number of averaged simulations. The ASE curves are compared to the result for a polarization-modulated repumper laser (modulation frequency $0.3\Gamma$ and amplitude $\pi$, see Ref.\ \cite{Lindvall2012a} for details) [dotted (blue) curves] and for a single-mode repumper laser with stationary polarization [dashed (red) curves]. In both cases the Rabi frequency was $\Omega_{\mathrm{r},\text{laser}}/\Gamma=1$, as discussed above, and the detuning was $\delta_{\mathrm{r},\text{laser}}/\Gamma=0.5$ in order to tune the system away from the coherent population trapping (CPT) resonances that otherwise can occur between the ground and metastable states. For the single-mode laser with stationary polarization, the scattering rate decays exponentially at the optical pumping rate $\Gamma_\text{OP}/\Gamma = 0.002$, whereas the scattering rate for the modulated laser stays constant after a small initial decay except for oscillations at the modulation frequency (an extended plot up to $\Gamma t = 2000$, comparing a coherent repumper with fixed polarization to one with modulated polarization, can be found in Fig.~8 of Ref.\ \cite{Lindvall2012a}).

As can be seen from Figs.~\ref{fig:t-plot}(a)--\ref{fig:t-plot}(c), the differences between the results for the three field models are negligible. Except for the residual fluctuations, the three curves are also very close to the three-level solution that will be derived in Sec.~\ref{sec:3-level}, shown in Fig.~\ref{fig:t-plot}(d) [solid (black) curve].

In practice, one is mainly interested in the time-averaged quasi-steady-state scattering rate, i.e., when the only time dependence is due to the fluctuations. For the parameters in Fig.~\ref{fig:t-plot}, both the PD model and the ASE model with different values of $g^{(2)}_{\text{ASE},ij}(0)$ (i.e., including the chaotic case) yield a scattering rate of $(7.8\pm 0.2) \times 10^{6}\;\mathrm{s}^{-1}$ for $^{88}\mathrm{Sr}^+$ when averaged over 10 simulations, with the variations between single simulations being larger than the variation between the averages for the different models. The corresponding scattering rate for the modulated laser is $8.7\times 10^{6}\;\mathrm{s}^{-1}$, i.e., the scattering rate for the ASE source is only about 10\% lower.
For a nonmodulated laser repumper, the scattering rate drops to $80\;\mathrm{s}^{-1}$, i.e., practically to zero. This value can also be estimated as the scattering rate obtained for a modulated laser multiplied by the ratio $\gamma_m/\Gamma_\text{op}$, where $\gamma_m = 2.3\;\mathrm{s}^{-1}$ is the decay rate of the metastable state and $\Gamma_\text{OP}\approx 0.002\Gamma$ is the rate of optical pumping into this state \cite{Lindvall2012a}.

The effects of a PD field and a chaotic field on a two-level atom have been compared by Georges and Lambropoulos \cite{Georges1979a}. They conclude that the difference is small when the saturation parameter $S = 2 C_\text{rms}^2\Omega_\mathrm{r}^2 / \Gamma (\Gamma + b) \ll 1$. Here, $C_\text{rms}$ is the root-mean-square transition amplitude that will be derived in Sec.~\ref{sec:3-level}. For the parameters in Fig.~\ref{fig:t-plot}, $S=1/3$, but this system is much more complex than a two-level atom: the repumper is driving one leg of a $\Lambda$ system with Zeeman substructure and the main mechanism that reduces the scattering rate is the buildup of coherent dark states between the sublevels, not saturation. Our results indicate that the amount of amplitude fluctuations and the exact statistics of the field are not relevant for the ASE repumper to work.

\begin{figure}[t]
\includegraphics[width=.85\columnwidth]{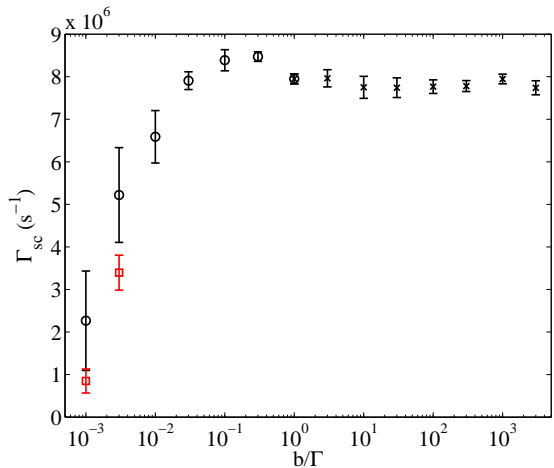}%
\caption{\label{fig:b-dep} (Color online) Time-averaged quasi-steady-state scattering rate as a function of repumper bandwidth $b/\Gamma$ for an ASE repumper with $g^{(2)}_{\text{ASE},ij}(0) = 1.5$. For $b/\Gamma\leq 1$, the repumper Rabi frequency was $\Omega_\mathrm{r}/\Gamma=1$ (black circles), whereas for $b/\Gamma\geq 1$, it was $\Omega_\mathrm{r}/\Gamma=\sqrt{b/\Gamma}$ (black crosses). For the smallest bandwidths, the corresponding values for the PD model are shown for comparison [gray (red) squares]. The circles, crosses, and squares are the mean and the error bars the standard deviation of ten simulations. $\delta_\mathrm{r}/\Gamma=0.5$.}
\end{figure}

Figure~\ref{fig:t-plot} was plotted for a bandwidth of $b/\Gamma=100$. However, the unpolarized repumper principle works over a wide range of bandwidths. Figure~\ref{fig:b-dep} shows the time-averaged quasi-steady-state scattering rate as a function of the bandwidth for an ASE repumper with $g^{(2)}_{\text{ASE},ij}(0) = 1.5$ (black circles and crosses). It shows that, for small bandwidths, the polarization fluctuations are too slow to efficiently destabilize dark states, so the scattering rate is low and the differences between individual simulations are large. Also shown in Fig.~\ref{fig:b-dep} is the corresponding data for the PD model for the smallest bandwidths [gray (red) squares]. Here, the scattering rate is slightly lower and the simulation-to-simulation variations smaller than for the ASE model, presumably due to the fact that the phase fluctuates much more smoothly in the PD model than in the ASE and the chaotic models. On the other hand, varying $g^{(2)}_{\text{ASE},ij}(0)$ from $1$ to $2$ in Eq.~(\ref{eq:E_ASE}) does not affect the results of the ASE model significantly, indicating that the amplitude fluctuations are of minor importance for the destabilization of dark states. For bandwidths $b/\Gamma\gtrsim 0.01$, there are no significant differences between the scattering rates for the different models.

For $b/\Gamma \approx 0.1 \ldots 0.3$, dark states are effectively destabilized while the bandwidth is still narrow, and thus the scattering rate is approximately the same as for a polarization-modulated laser.
As $b/\Gamma$ is increased above 1, with the Rabi frequency scaling as $\Omega_\mathrm{r}/\Gamma= (b/\Gamma)^{1/2}$, the RMS fluctuations of the scattering rate (in a single simulation) increase, but at $b/\Gamma\approx 10$ the fluctuations have essentially reached their maximum amplitude.
More importantly, the time-averaged quasi-steady-state scattering rate remains essentially constant for $b/\Gamma \geq 1$ (see Fig.~\ref{fig:b-dep}). Thus for $b/\Gamma \gg 1$ the time-averaged quasi-steady-state scattering rate depends only on the ratio $\Omega_\mathrm{r}^2/b$, proportional to the PSD at line center. The bandwidths of real ASE sources can be of the order of $1\ldots 10$\;nm, corresponding to $b/\Gamma \approx 10^4\ldots 10^5$, but simulating this large bandwidths is not practically feasible.

It should be noted that in a single-ion experiment, the fluctuations in the scattering rate due to the stochastic ASE field will not be detected, since the scattered photon detection time constant typically is several orders of magnitude greater than the time scale of the fluctuations. In fact, the time constant is longer than all time scales of the ion system and thus one is essentially measuring the ensemble-averaged quasi-steady-state scattering rate.

\section{Partially polarized field \label{sec:PP field}}

In order for the ASE repumper to be a reliable, practical device, it is important that it works even if the output light is not completely unpolarized. The ASE field (\ref{eq:unpol}) can be made partially polarized by redefining the electric field of the second polarization component $\mathbf{u}_2$ as
\begin{equation}
\mathcal{E}_{2,\text{PP}}(t) = c_1 \mathcal{E}_1(t) + c_2 \mathcal{E}_2(t),
\end{equation}
where $c_1$ and $c_2$ are complex numbers obeying $|c_1|^2+|c_2|^2 = 1$. The degree of polarization \cite{Al-Qasimi2007a} of the total field is then given by $P=|c_1|$. If $c_1$ is real, the polarized part is linear, if it is purely imaginary, it is circular.

\begin{figure}[h]
\includegraphics[width=.85\columnwidth]{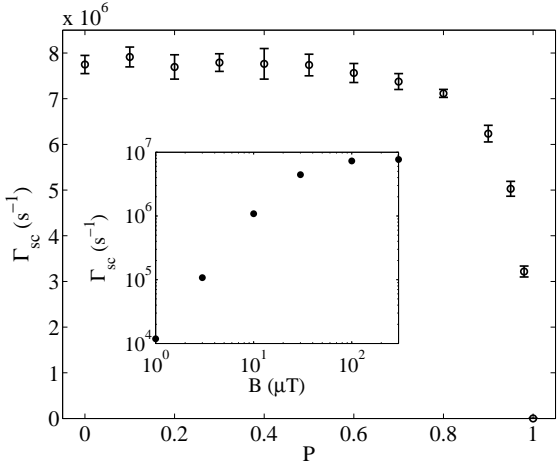}%
\caption{\label{fig:Pdep}Time-averaged quasi-steady-state scattering rate as a function of the degree of polarization for an ASE field with $g^{(2)}_{\text{ASE},ij}(0)=1.5$ and $b/\Gamma=100$. The circles are the mean and the error bars the standard deviation of ten simulations. The inset shows the scattering rate as a function of the external magnetic field $B$ for $P=1$.}
\end{figure}

Figure~\ref{fig:Pdep} shows the time-averaged quasi-steady-state scattering rate as a function of $P$. The ASE model with $g^{(2)}_{\text{ASE},ij}(0)=1.5$ and $b/\Gamma=100$, and a linear partial polarization were used for these simulations, but we have found no significant differences for the other field models, other values of $g^{(2)}_{\text{ASE},ij}(0)$, other bandwidths, or a circular partial polarization.
The scattering rate remains within 10\% of the maximum value up to $P\approx 0.8$, demonstrating that the ASE repumper is not sensitive to the output being partially polarized. For fully polarized light, $P=1$, we obtain the same scattering rate $\Gamma_\text{sc} = 80\;\mathrm{s}^{-1}$ as for an unmodulated laser. The inset in Fig.~\ref{fig:Pdep} shows the scattering rate as a function of the external magnetic field for $P=1$. We note that if the ASE light is completely polarized, the magnetic field required to destabilize dark states is of the same magnitude as for a laser repumper \cite{Lindvall2012a}, regardless of the fact that the ASE is incoherent and broadband.

\section{Reduced three-level system \label{sec:3-level}}

The numerical eight-level simulations described in Sec.~\ref{sec:8-level} are very computer-time consuming. We are therefore motivated to derive a simpler, approximative system that can be used, e.g., to find the optimum power spectral density (PSD) of the ASE repumper for different cooling-laser Rabi frequencies and detunings.

The purpose of the ASE repumper is to provide efficient repumping to the cooling cycle while preventing coherences from building up between sublevels of the metastable state and between sublevels of the ground and metastable states. The numerical simulations in Sec.~\ref{sec:8-level} show that these coherences fluctuate around zero with small amplitudes. Hence they vanish in an ensemble average and we set the metastable coherences $\rho_{mm'}$ and the ground-metastable coherences $\rho_{gm}$ to zero in order to simplify the system. Also justified by the numerical results, we set the excited state coherence $\rho_{78}$ to zero, as the ASE repumper is incoherent and the linearly polarized cooling laser only drives the $|1\rangle \rightarrow |7\rangle$ and $|2\rangle \rightarrow |8\rangle$ transitions without coupling the two excited-state levels to each other. With the previous simplifications, one notices that the density matrix elements $\rho_{18}$, $\rho_{27}$, and $\rho_{12}$ only depend on each other. As they initially are zero, they will remain so and can be discarded.

The $m$-$e$ optical coherences decay at the rate $b/2 \gg \Gamma$. We can therefore use the ``broad-line approximation'' \cite{Ducloy1973a,Blushs2004a}, which assumes that the optical coherences follow the populations adiabatically so that we can use the steady-state expressions obtained by setting $d\rho_{me}/dt = 0$,
\begin{equation} \label{eq:m-e_cohSS}
\rho_{me} = -i \frac{\Omega_\mathrm{r} f_q^* C_{em}}{b} (\rho_{mm}-\rho_{ee}),
\end{equation}
where $C_{em}$ is the relative transition amplitude, $f_q^*$ is the complex conjugate of the polarization component amplitude, and $q=0$, $\pm1$ is the polarization that connects the two $|m\rangle$ and $|e\rangle$ sublevels (see Ref.\ \cite{Lindvall2012a} for details). Substituting Eq.~(\ref{eq:m-e_cohSS}) into the density-matrix equations, we are left with ten equations: the eight populations and the two $g$-$e$ optical coherences $\rho_{17}$ and $\rho_{28}$.

Because of symmetry, we must have $\rho_{11}=\rho_{22}$ and $\rho_{77}=\rho_{88}$. As $C_{71}=-C_{82}$ \cite{Lindvall2012a} and the two transitions $|1\rangle \rightarrow |7\rangle$ and $|2\rangle \rightarrow |8\rangle$ are driven by the same polarization component, it follows from Eq.~(\ref{eq:m-e_cohSS}) that  $\rho_{17}=-\rho_{28}$. In addition, from the numerical simulations we know that the metastable-level populations are approximately equal. We can thus define the total populations $\rho_{gg}=\rho_{11}+\rho_{22}$, $\rho_{mm}=\sum_{k=3}^6 \rho_{kk}$, and $\rho_{ee}=\rho_{77}+\rho_{88}$. If we furthermore formally define $\rho_{ge}=2\rho_{28}=\rho_{28}-\rho_{17}$, we obtain the equations for the reduced three-level system [in the rotating wave approximation and rotating frame, see Fig.~\ref{fig:8_3-level}(b)],
\begin{subequations} \label{eq:3-level}
\begin{eqnarray}
\dot\rho_{ee} &=& -\Gamma\rho_{ee} +\frac{C_\text{rms}^2\Omega_\mathrm{r}^2}{b}\left(\frac{\rho_{mm}}{2}-\rho_{ee}\right) -C_{82}\Omega_\mathrm{c}\rho_{ge}^\mathrm{i}, \\
\dot\rho_{mm} &=& A_m\Gamma\rho_{ee} -\gamma_m\rho_{mm} -\frac{C_\text{rms}^2\Omega_\mathrm{r}^2}{b}\left(\frac{\rho_{mm}}{2}-\rho_{ee}\right), \\
\dot\rho_{gg} &=& A_g\Gamma\rho_{ee}+\gamma_m\rho_{mm}+C_{82}\Omega_\mathrm{c}\rho_{ge}^\mathrm{i}, \\
\dot\rho_{ge} &=& -\left( \frac{\Gamma}{2} +i\delta_\mathrm{c} \right) \rho_{ge} + i\frac{C_{82}\Omega_\mathrm{c}}{2} (\rho_{ee}-\rho_{gg}).
\end{eqnarray}
\end{subequations}
Here, $\rho_{ge}^\mathrm{i} = \mathrm{Im}(\rho_{ge})$, $A_g$ and $A_m$ are the decay probabilities to the states $|g\rangle$ and $|m\rangle$, respectively ($A_g + A_m=1$), $\gamma_m$ is the decay rate of the metastable state, and $\delta_\mathrm{c}$ is the cooling laser detuning. The root-mean-square transition amplitude $C_\text{rms}$ is obtained from $C_\text{rms}^2=|f_{-1}^\mathrm{r}C_{86}|^2 +|f_{0}^\mathrm{r}C_{85}|^2 +|f_{+1}^\mathrm{r}C_{84}|^2 = |f_{-1}^\mathrm{r}C_{75}|^2 +|f_{0}^\mathrm{r}C_{74}|^2 +|f_{+1}^\mathrm{r}C_{73}|^2 = 1/6$, where the average polarization probabilities for an unpolarized plane wave are $|f_{-1}^\mathrm{r}|^2=|f_{+1}^\mathrm{r}|^2=1/4$ and $|f_{0}^\mathrm{r}|^2=1/2$.

Note that Eqs.~(\ref{eq:3-level}) depend only on the ratio $\Omega_\mathrm{r}^2/b$, which was already observed for the eight-level results in Sec.~\ref{sec:8-level}.
These equations can be numerically integrated to obtain the time-dependent solution and Fig.~\ref{fig:t-plot} shows that the solution agrees very well with the numerical eight-level results.

The steady-state solution of Eqs.~\ref{eq:3-level} can be solved exactly to give the excited state population
\begin{widetext}
\begin{equation} \label{eq:rho_ee}
\rho_{ee} = \frac{\left(\frac{C_{82}\Omega_\mathrm{c}}{2}\right)^2} {\delta_\mathrm{c}^2 +\left(\frac{\Gamma}{2}\right)^2 +\left(\frac{C_{82}\Omega_\mathrm{c}}{2}\right)^2 + \left(\frac{R}{2}+\gamma_m\right)^{-1} \left\{ \left(\frac{C_{82}\Omega_\mathrm{c}}{2}\right)^2 \left[ \frac{3}{2}R +A_m\Gamma +\gamma_m\right] -R\left(\frac{A_m}{2}-\frac{\gamma_m}{\Gamma}\right) \left[\delta_\mathrm{c}^2 +\left(\frac{\Gamma}{2}\right)^2\right] \right\}},
\end{equation}
\end{widetext}
where $R=C_\text{rms}^2\Omega_\mathrm{r}^2/b$. Since the metastable state decay rate $\gamma_m$ is much smaller than both $\Gamma$ and $R$, Eq.~(\ref{eq:rho_ee}) can further be simplified to
\begin{equation} \label{eq:rho_ee-a}
\rho_{ee} \approx \frac{\left(\frac{C_{82}\Omega_\mathrm{c}}{2}\right)^2} {A_g\left[\delta_\mathrm{c}^2 +\left(\frac{\Gamma}{2}\right)^2\right] +\left(\frac{C_{82}\Omega_\mathrm{c}}{2}\right)^2 \left( 4+2A_m\frac{\Gamma}{R}\right)}.
\end{equation}
The last term in the denominator of Eq.~(\ref{eq:rho_ee-a}) describes optical pumping and is similar to what one obtains if the populations of the $|g\rangle$ and $|m\rangle$ sublevels are equalized by a relaxation process such as collisions \cite{Lindvall2007a}.
Note, however, that the exact solution (\ref{eq:rho_ee}) simplifies to the familiar expression for a two-level system in the limit $A_g \rightarrow 1$ ($A_m \rightarrow 0$) and $R \rightarrow 0$, which the approximation (\ref{eq:rho_ee-a}) does not.

\begin{figure}[h]
\includegraphics[width=.95\columnwidth]{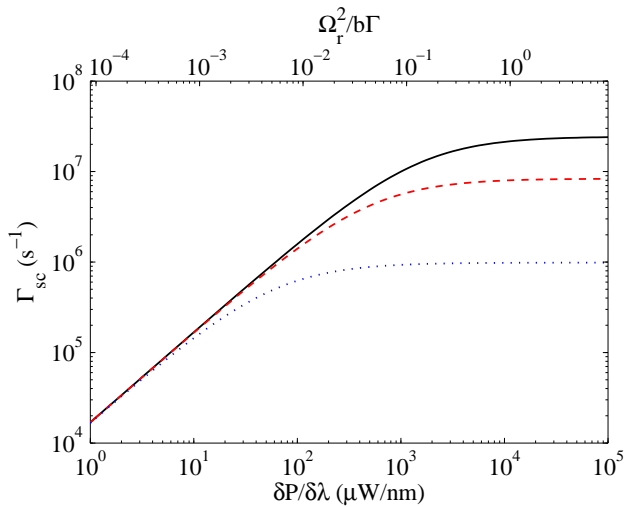}%
\caption{\label{fig:ASE-scatt}(Color online) Scattering rate as a function of $\Omega_\mathrm{r}^2/(b\Gamma)$ (top axis) and PSD (bottom axis, assuming a beam waist $w=50\;\mu$m) for three different cooling laser Rabi frequencies: $\Omega_\mathrm{c}/\Gamma=3$ [solid (black) curve], $\Omega_\mathrm{c}/\Gamma=1$ [dashed (red) curve], and $\Omega_\mathrm{c}/\Gamma=0.3$ [dotted (blue) curve].}
\end{figure}

For the parameters used in Fig.~\ref{fig:t-plot}, the three-level steady-state scattering rate $\Gamma_{sc} = A_g \Gamma \rho_{ee}$ is $8.0\times 10^{6}\;\mathrm{s}^{-1}$, in good agreement with the numerical result of $(7.8\pm 0.2) \times 10^{6}\;\mathrm{s}^{-1}$ obtained in Sec.~\ref{sec:8-level}. Figure~\ref{fig:ASE-scatt} shows the scattering rate as a function of both $\Omega_\mathrm{r}^2/(b\Gamma)$ and the experimentally more accessible PSD $\delta P/\delta \lambda$ (assuming a beam waist $w=50\;\mu$m) for three different cooling laser Rabi frequencies. The figure shows that for each value of $\Omega_\mathrm{c}$, there is a limiting ASE PSD, above which the scattering rate does not increase anymore. As anticipated in Sec.~\ref{sec:8-level}, this value is $\Omega_\mathrm{r}^2/b \approx \Omega_\mathrm{c}^2/\Gamma$. However, note that one comes very close to the maximum scattering rate already for PSDs one order of magnitude lower than this value.

\section{Experimental considerations \label{sec:experimental}}

A sufficient PSD at the Sr$^+$ repumping wavelength of 1092\;nm can be obtained from the ASE of a laser-pumped ytterbium-doped fiber \cite{FordellTBS}. This ASE source consists of a pump diode laser and standard fiber optic components. The fact that it requires no frequency stabilization or external polarization modulation not only provides greater simplicity and reliability in operation, but also makes it a cost-effective alternative to, e.g., a custom-ordered distributed feedback (DFB) laser.
The emitted light is unpolarized and its transversal coherence is similar to that of a laser, so it can be focused to the spot size required for an ion trap.

Current SLEDs can deliver a few hundred $\mu\mathrm{W}/\mathrm{nm}$ at 1092\;nm. As these typically emit nearly fully polarized light, two SLEDs have to be combined with orthogonal polarizations. Taking into account losses and the need for filtering (see below), this PSD is at the lower limit of efficient ASE repumping. However, more powerful SLEDs are expected to become available in the future.

One important question that needs to be addressed is whether there are any drawbacks related to the high total intensity of the ASE field. If one uses a bandpass filter at the cooling-laser wavelength for fluorescence detection and ion imaging, these will not be affected by the repumper light. The ASE spectrum might have a broad pedestal around the peak at 1092\;nm. If this is the case, light at the clearout wavelength 1033\;nm ($^2D_{5/2}$-$\,^2P_{3/2}$ transition) has to be filtered out, otherwise the repumper interferes with the detection of clock transitions.

Due to the high total intensity of the ASE light, the light shift of the reference transition is large. If we assume an ASE spectral width of 2\;nm (FWHM), corresponding to $b/\Gamma \approx 2.3\times 10^4$, and require $\Omega_\mathrm{r}/\Gamma= (b/\Gamma)^{1/2}\approx 150$, the light shift is $-14$\;kHz. The ASE repumper light must therefore be carefully blocked during the clock pulse. The ASE source can be switched off electronically by turning off the pump (laser) current in order to further increase the extinction ratio of the mechanical shutter or acousto-optic modulator (AOM) used for switching.
For the ASE source parameters above, the light shifts of the cooling and clearout transitions are $4$ and 36\;kHz, respectively, and can be neglected compared to the natural linewidth of these electric dipole  transitions.

The transition strength of the clearout transition (see Fig.~\ref{fig:energy-levels}) is similar to that of the repumping transition, and an ASE source could be used also to replace the clearout laser that is used to return the ion to the cooling cycle after a quadrupole transition has occurred. In this case, the main advantage is that no frequency stabilization is required.

\section{Application to other ions \label{sec:ions}}

If we first consider ions without hyperfine structure, the scheme described for $^{88}\mathrm{Sr}^+$ in Sec.~\ref{sec:introduction} applies also to the commonly used alkaline-earth-metal ions $^{40}\mathrm{Ca}^+$ \cite{Chwalla2009a} (repumper wavelength 866\;nm) and $^{138}\mathrm{Ba}^+$ \cite{Kurz2010a} (650\;nm), as well as to other less abundant even isotopes of these elements. Also in these ions, the clock transition is the $^2S_{1/2}$-$\,^2D_{5/2}$ quadrupole transition. The $\mathrm{Yb}^+$ ion has a similar energy level scheme, except that its lowest excited state is the extremely metastable $^2F_{7/2}$ state. For $\mathrm{Yb}^+$, the upper state of the repumping transition is usually the $^3D[3/2]_{1/2}$ state \cite{Bell1991a} (see Fig.~\ref{fig:energy-levels}) and an ASE repumper at 935\;nm could be used for $^{172}\mathrm{Yb}^+$ both for the $^2S_{1/2}$-$\,^2D_{5/2}$ quadrupole clock transition \cite{Taylor1997a} and the $^2S_{1/2}$-$\,^2F_{7/2}$ octupole clock transition \cite{Roberts1997a}.

Odd isotopes with nuclear spin $I$, thus featuring hyperfine structure, also have dark states within the $^2D_{3/2}$ state. In $^{87}\mathrm{Sr}^+$ with $I=9/2$, there are dark states in the $F=5$ and $F=6$ hyperfine levels that require polarization modulation \cite{Boshier2000a}. In addition, due to the hyperfine structure, there are three repumping transitions that need to be driven. This has been solved by frequency modulating the repumping laser over the few hundred MHz range of the hyperfine structure \cite{Barwood2003a}. In addition to destabilizing the dark states, a broadband ASE repumper would solve this problem by driving all allowed $^2D_{3/2}$-$\,^2P_{1/2}$ hyperfine transitions.

Of the odd isotopes, $^{171}\mathrm{Yb}^+$ is particularly promising for optical clocks, as it has a minimal hyperfine structure due to its nuclear spin of $1/2$ and, in addition, has three possible clock transitions. Here, an ASE repumper could be used with the $^2S_{1/2}$-$\,^2D_{5/2}$ quadrupole and $^2S_{1/2}$-$\,^2F_{7/2}$ octupole \cite{King2012a,Huntemann2012a} clock transitions. At the National Physical Laboratory (NPL), a magnetic field around $20\;\mu$T is used to destabilize dark states in the $^2D_{3/2}$ manifold and two separate repumper lasers are used to empty its two hyperfine levels $F=1$ and $F=2$ \cite{King2012a}. At Physikalisch-Technische Bundesanstalt (PTB), a high laser intensity is used in order to deplete both the hyperfine levels and a 1 mT magnetic field, applied during the cooling period, destabilizes dark states in both the $^2D_{3/2}$ manifold and the $^2S_{1/2}(F=1)$ state \cite{Tamm2009a}. Again, a broadband ASE repumper could drive all the allowed repumping transitions without creating dark states in arbitrarily low magnetic fields. An ASE repumper cannot be directly utilized with the third clock transition, $^2S_{1/2}(F=0)$\,-$\,^2D_{3/2}(F=2)$, as the repumper in this case must be able to selectively address the $^2D_{3/2}(F=1)$ level in order for the electron shelving technique to work \cite{Tamm2009a}.

The oscillator strengths of the repumping transitions in $\mathrm{Ca}^+$ \cite{Safronova2011a} and $\mathrm{Ba}^+$ \cite{Gallagher1967a} are comparable to that in Sr$^+$ \cite{JiangD2009a}, so the required PSDs are also similar. The oscillator strength of the $^2D_{3/2}$-$\,^3D[3/2]_{1/2}$ transition in $\mathrm{Yb}^+$ is one order of magnitude lower \cite{Fawcett1991a,Biemont1998a}, and thus the required PSD is an order of magnitude higher. Naturally, the availability of suitable ASE sources has to be investigated for each wavelength.

\section{Conclusions}

We have proposed using unpolarized, incoherent amplified spontaneous emission (ASE) to drive the repumping transition in a trapped and laser-cooled single ion. We have constructed a theoretical model for the ASE radiation and analyzed its performance compared to a single-mode laser repumper by solving the eight-level density-matrix equations for a Sr$^+$ ion. The performance of the ASE repumper depends mainly on its power spectral density at the repumping wavelength and it works even if the field is partly polarized. We have also derived a reduced three-level system that can be solved exactly, enabling rapid comparison with experiments and optimization of experimental parameters. The required ASE power spectral density can be obtained with current technology and the idea can be applied also to the clearout transition and to several other commonly used ions.

An ASE repumper prevents dark states from forming without external polarization modulation even in zero magnetic field. In addition, it requires no frequency stabilization due to the broad bandwidth. This makes it robust and compact, which is particularly important for transportable ion clocks.


\begin{acknowledgments}
The authors would like to thank Dr.\ A.~A.~Madej for carefully reading and commenting the manuscript. This work was funded by the Academy of Finland (Project No.\ 138894). T.F. gratefully acknowledges financial support from the European Commission (Marie Curie Integration Grant PCIG10-GA-2011-304084).
\end{acknowledgments}


%

\end{document}